\documentclass[aps,prb, twocolumn,showpacs,amsmath,amssymb,superscriptaddress,floatfix]{revtex4-2}

\usepackage{color}        
\usepackage[pdftex]{graphicx}     
\usepackage{bm}
\usepackage{natbib}            
\usepackage[colorlinks,plainpages=false,linkcolor=black,urlcolor=black,citecolor=black,pdfpagemode=UseNone,pdfstartview=FitBH]{hyperref}
\usepackage{float}
\usepackage{todonotes}

\begin{document}

\title{Fermi surface geometry and momentum dependent electron-phonon coupling \\ drive the charge density wave in quasi-1D ZrTe$_3$}

\author{Josu Diego}
\email{josu.diegolopez@unitn.it}
\affiliation{Dipartimento di Fisica, Università di Trento, Via Sommarive 14, 38123 Povo, Italy.}

\author{Matteo Calandra}
\email{m.calandrabuonaura@unitn.it}
\affiliation{Dipartimento di Fisica, Università di Trento, Via Sommarive 14, 38123 Povo, Italy.}

\date{\today}
\begin{abstract}

ZrTe$_3$ is a prototypical quasi-one-dimensional compound undergoing a charge density wave  transition via a very sharp Kohn anomaly in phonon momentum space. While Fermi surface geometry has long been considered the primary driver of the instability, a full understanding of the lattice dynamics and electron-phonon role has remained elusive. Our first principles calculations in the high-symmetry phase show that the Fermi surface is correctly reproduced only when the Hubbard interaction on the Te $5p$ orbitals is included, which in turn is essential for the appearance of a soft harmonic phonon mode at the CDW wavevector. Analyzing the mode and momentum dependence of the electron-phonon coupling, we find that its variations with phonon momentum dominate over electronic effects. These results identify unambiguously the CDW origin in ZrTe$_3$ as a cooperative effect of Fermi surface geometry and momentum-dependent electron-phonon coupling, with the latter playing the leading role. We further determine the atomic structure in the low-symmetry CDW phase, revealing a nonchiral modulation. The mechanisms revealed in our work are directly relevant to other quasi-1D systems, including trichalcogenides and compounds hosting Peierls-like chains.

\end{abstract}

\maketitle

\section{Introduction}

Charge density waves (CDWs) in low-dimensional materials are often interpreted within the Peierls framework \cite{peierls1955}, where a one-dimensional (1D) metallic chain with a partially filled band becomes unstable and spontaneously develops a periodic modulation of the electronic density. In real crystals, compounds with chainlike structures provide the closest analogues to an ideal Peierls chain, as certain electronic states propagate mainly along these atomic wires and generate quasi-1D Fermi surfaces prone to electronic instabilities. Prototypical examples of quasi-1D CDW materials include platinum chain compounds \cite{KCP}, the blue bronzes \cite{Pouget1983-ws, Pouget1989}, and the transition metal trichalcogenides \cite{nbse3,tas3}, where nearly parallel Fermi surface sheets can be connected by a characteristic wavevector matching the CDW periodicity. Since the Fermi surface geometry and the consequent electronic instability offer a natural explanation for the selection of the CDW wavevector, the role of electron-phonon coupling in stabilizing the ordered phase is frequently overlooked in theoretical analyses and therefore remains poorly understood, especially in these quasi-1D systems where nesting is strongest. Here, we focus on ZrTe$_3$, a transition metal trichalcogenide, as a model system to explore the interplay between electronic instabilities and lattice dynamics in the CDW formation. \\

\begin{figure}[t]
\includegraphics[width=\columnwidth]{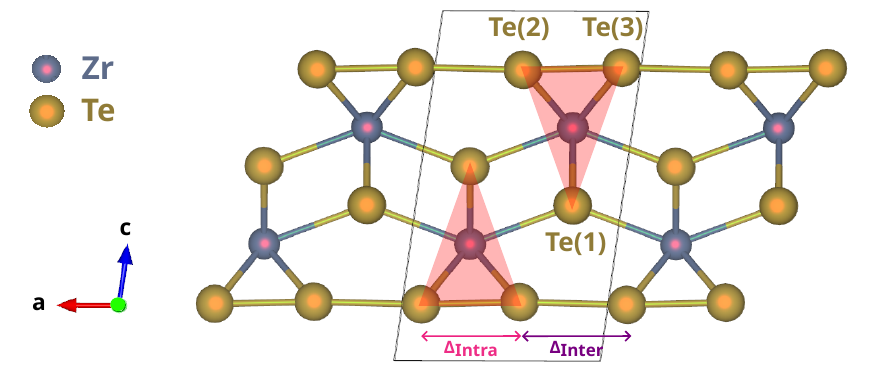}
\caption{Crystal structure of ZrTe$_3$ viewed along the \textbf{b}-axis. The unit cell is outlined with black lines, while shaded areas inside indicate the Zr-centered prisms along \textbf{b}. Te(2) and Te(3) atoms forming the secondary chain along \textbf{a} are highlighted. Alternating intraprismatic ($\Delta_{\rm Intra}$) and interprismatic ($\Delta_{\rm Inter}$) distances along this dimerized chain are listed in Table~\ref{tab:Te_distances}.}
\label{fig: crystal}
\end{figure}

ZrTe$_3$ undergoes a charge density wave transition at T$_{\text{CDW}}$ = 63 K, which was first identified as an anomalous hump in the resistivity along both \textbf{a} and \textbf{c} crystallographic directions  \cite{Takahashi1983}. Despite the onset of charge modulation, the material remains metallic, and filamentary SC along \textbf{a} emerges at T$_{\text{c}}$ = 2 K \cite{Nakajima1986}. Electron-diffraction measurements revealed that the associated periodic lattice distortion is defined by a modulation vector of \textbf{q}$_{\text{CDW}}$ = (0.07, 0, 0.33) r.l.u. (reciprocal lattice units) \cite{Eaglesham1984}. This wavevector shows no component along the prismatic chains, in sharp contrast to the majority of quasi-1D transition metal trichalcogenides where the CDW develops along the chain direction \cite{Monceau2012}. This peculiarity of ZrTe$_3$ arises from the strong interchain Te(2)-Te(3) bonds that create an additional chainlike motif along \textbf{a}, to which the CDW is primarily linked (see Fig. \ref{fig: crystal}). Indeed, electronic structure calculations \cite{Stowe1998, Felser1998} and ARPES experiments \cite{Yokoya2005, Starowicz2007} reveal that the high-temperature Fermi surface contains a pair of quasi-1D, nearly flat horizontal sheets (see Fig. \ref{Fig1}) originated from Te(2)-Te(3) $5p_x$ bands that are nested by \textbf{q}$_{\text{CDW}}$. ARPES studies further show that the CDW gap opens selectively on the perfectly nested Fermi surface sections, in particular near the D point of the Brillouin zone \cite{Yokoya2005, Hoesch2019}. Altogether, the experimental and theoretical characterization of the electronic structure of ZrTe$_3$ associate the CDW to a Fermi surface mechanism. From the vibrational point of view, the transition is accompanied by a soft phonon mode whose frequency decreases sharply over a narrow momentum range upon approaching T$_{\rm CDW}$ from above, ultimately freezing into the static lattice modulation below the critical temperature~\cite{Hoe09}. Such a strongly localized softening in momentum space is likewise considered a fingerprint of an electronic instability driven by the Fermi surface geometry. \\

Despite the extensive electronic structure investigations of ZrTe$_3$, previous works have focused on Fermi surface nesting as the relevant criterion in the CDW formation, overlooking that an electronically driven transition is signaled by a significant enhancement of the real part of the electronic susceptibility \cite{johannes2009}. Moreover, first principles studies of its vibrational properties and of the role played by the electron-phonon interaction driving the ordered state remain surprisingly scarce. To the best of our knowledge, the only theoretical-experimental work addressing this aspect has focused on the analysis of anomalously broad Raman peaks, highlighting strong electron-phonon coupling effects but providing information limited to $\Gamma$-point phonons \cite{eph2015}. A full understanding of the microscopic origin of the charge modulated state, however, requires access to the lattice dynamics at the CDW wavevector itself. Only an explicit calculation of the phonon spectrum and of the momentum dependence of the electron-phonon interaction can clarify the mechanism driving the transition. \\

In this work we address this problem by presenting a comprehensive \textit{ab initio} study of the electronic structure, lattice dynamics, and electron-phonon interaction of the high-temperature (HT) phase of ZrTe$_3$. We show that an accurate description of the Fermi surface requires the inclusion of correlation effects on the Te $5p$ orbitals. Only at this level of theory a soft harmonic phonon mode emerges at the experimental CDW wavevector. By analyzing the momentum and mode dependence of the electron-phonon coupling, we demonstrate that, while the CDW is partly due to a Fermi surface instability, the strong variations of the electron-phonon interaction in momentum space actively drive the lattice distortion. Finally, we determine the atomic structure of the low-symmetry CDW phase, which has not been fully resolved experimentally and has been proposed to be chiral.

\section{Computational details}

All calculations were performed in the high-symmetry phase of ZrTe$_3$,  employing the experimental lattice parameters measured at room temperature and taken from Ref. \cite{Stowe1998}. Electronic and vibrational properties were calculated from first principles with the {\sc Quantum Espresso} package~\cite{0953-8984-21-39-395502,0953-8984-29-46-465901}. The electronic structure was computed based on plane-wave DFT within the Perdew-Burke-Ernzerhof (PBE)~\cite{PBE1996}  exchange-correlation functional. We used ultrasoft pseudopotentials \cite{Garrity2014} including 4$s^{2}$ 4$p^6$ 4$d^2$ 5$s^2$ electrons in the valence for Zr, and 5$s^2$ 5$p^4$ for Te. The plane-wave basis set was truncated at a kinetic energy cutoff of 35 Ry for the wavefunctions and 350 Ry for the charge density. In addition, DFT+U calculations were performed to include electron-electron interactions on the Te $5p$ states. The on-site Hubbard $U$ parameters were computed from first principles using the linear-response method of Timrov et al. \cite{Timrov1, Timrov2}, which yields screened values defined in terms of bare and fully screened susceptibilities \cite{Carta2025}. The projectors were built from the Te $5p$ atomic orbitals in the pseudopotential and employing a $3\times3\times3$ \textbf{q}-point grid. We also tested the inclusion of a Hubbard $U$ on Zr $4d$ orbitals; however, it produces only negligible changes in the band structure (see Fig. S1 in the Supplemental Material \cite{supplemental_material}), and therefore is not considered further. Calculations including spin-orbit coupling (SOC) were also carried out using projector augmented wave (PAW) pseudopotentials~\cite{DalCorso2014} with the same valence electron configurations. In this case, the energy cutoffs were increased to 90 Ry for the wavefunctions and 360 Ry for the charge density. In all cases both structural relaxation (with the original crystal symmetry preserved) and electronic calculations were performed using a 20$\times$25$\times$10 \textbf{k}-point grid and a Methfessel-Paxton smearing \cite{PhysRevB.40.3616} of 0.001 Ry.\\

Phonon frequencies were calculated within the harmonic approximation using density functional perturbation theory (DFPT) \cite{dfpt} as implemented in {\sc Quantum ESPRESSO}, with DFPT+$U$ \cite{Floris1,Floris2} employed for correlated calculations. This approach allows for exact harmonic phonon calculations at each $\mathbf{q}$-point without the need for supercells. Since the Fermi surface of ZrTe$_3$ consists of two nearly parallel sheets at fixed $k_x$ (see Fig.~\ref{Fig1}), phonon convergence may require a dense sampling along this direction. We find that the frequency of the CDW driving phonon mode is converged using a $35 \times 15 \times 6$ \textbf{k}-point mesh and a Methfessel--Paxton smearing of 0.001 Ry (see Fig. S2 \cite{supplemental_material}). The electron-phonon linewidth $\gamma_{\mu}(\mathbf{q})$ was evaluated according to Eq.~(\ref{eqn: lw}), using a converged $100 \times 30 \times 12$ \textbf{k}-point grid and a Gaussian smearing of 0.001 Ry to approximate the electronic Dirac delta functions. \\

The nesting function, $\zeta(\mathbf{q})$, and the real part of the non--interacting electronic susceptibility, $\chi_0(\mathbf{q})$, were computed on a 200$\times$100$\times$60 \textbf{k}-point grid with the EPIq code \cite{MARINI2024108950}. Dense sampling of the Brillouin zone was achieved by constructing maximally localized Wannier functions (MLWFs) for entangled bands \cite{MLWF, MLWF2}. A total of 28 MLWFs were generated with the Wannier90 package \cite{Wan90} by projecting the initial Bloch states from a $15 \times 15 \times 6$ $\mathbf{k}$-point mesh onto Zr $d$ and Te $p$ orbitals. Furthermore, a broadening of 0.008 eV was applied to the Dirac delta functions in $\zeta(\mathbf{q})$.\\

The low-symmetry phase was obtained by displacing the atoms along the eigendisplacement of the unstable phonon mode at $\mathbf{q}_{\mathrm{CDW}}$ within a $14 \times 1 \times 3$ supercell nearly commensurate with the CDW wavevector. The minimum energy configuration along this direction was subsequently used as the starting point for a full structural relaxation. Both calculations were carried out using the same computational parameters as in the unit cell calculations, but with a reduced $5 \times 15 \times 2$ \textbf{k}-point grid. The density of states (DOS) of the relaxed structure was then computed using a denser $10 \times 25 \times 3$ \textbf{k}-point mesh.

\section{Results}

\subsection{Crystal Structure}

\begin{table}[t]
\centering
\renewcommand{\arraystretch}{1.5} 
\begin{tabular*}{\columnwidth}{@{\extracolsep{\fill}} lcc}
\hline
\hline
\textbf{Approach} & $\Delta_{\mathrm{Intra}}$ (\AA) & $\Delta_{\mathrm{Inter}}$ (\AA) \\
\hline
Exp. (293 K) \cite{Stowe1998} & 2.79 & 3.10 \\
PBE & 2.87 & 3.02 \\
PBE+SOC & 2.88 & 3.01 \\
PBE+U & 2.82 & 3.07 \\
\hline
\hline
\end{tabular*}
\caption{Intraprismatic ($\Delta_{\mathrm{Intra}}$) and interprismatic ($\Delta_{\mathrm{Inter}}$) Te(2)-Te(3) distances in the HT-phase of ZrTe$_3$ from experimental measurements and different electronic-structure approaches.}

\label{tab:Te_distances}
\end{table}

ZrTe$_3$ crystallizes in the monoclinic space group $P2_1/m$ (No. 11), with lattice parameters $a = 5.898$ Å, $b = 3.927$ Å, $c = 10.103$ Å, and $\beta = 97.81^\circ$ at room temperature \cite{Stowe1998}. All atoms occupy Wyckoff \textit{2e} positions, forming a quasi-two-dimensional layered structure along the \textbf{c}-axis. Each unit cell hosts two inversion-related ZrTe$_3$ prismatic chains running along the monoclinic \textbf{b}-axis (see Fig.~\ref{fig: crystal}), which attribute a marked quasi-one-dimensional character to the structure. Each chain is formed by Zr atoms coordinated by the three inequivalent Te sites Te(1), Te(2), and Te(3), with the chains interconnected through Zr-Te(1) bonds. Notably, the Te(2) and Te(3) atoms, located at the edges of the prisms, form a second dimerized chain along the \textbf{a}-axis, which is exposed to the interlayer region and plays a crucial role in the material’s electronic properties. The corresponding Te(2)-Te(3) distances are listed in Table~\ref{tab:Te_distances}, where $\Delta_{\rm Intra}$ and $\Delta_{\rm Inter}$ denote distances within a single prism and between adjacent ones, respectively, as obtained from experiment and from structural relaxations using different computational schemes. Among these approaches, PBE+U yields values in closest agreement with experimental measurements, highlighting the sensitivity of these chains to the treatment of electronic correlations. \\

\subsection{Electronic structure of the HT-phase}

\begin{figure*}
\begin{center}
\includegraphics[width=2.0\columnwidth,draft=false]{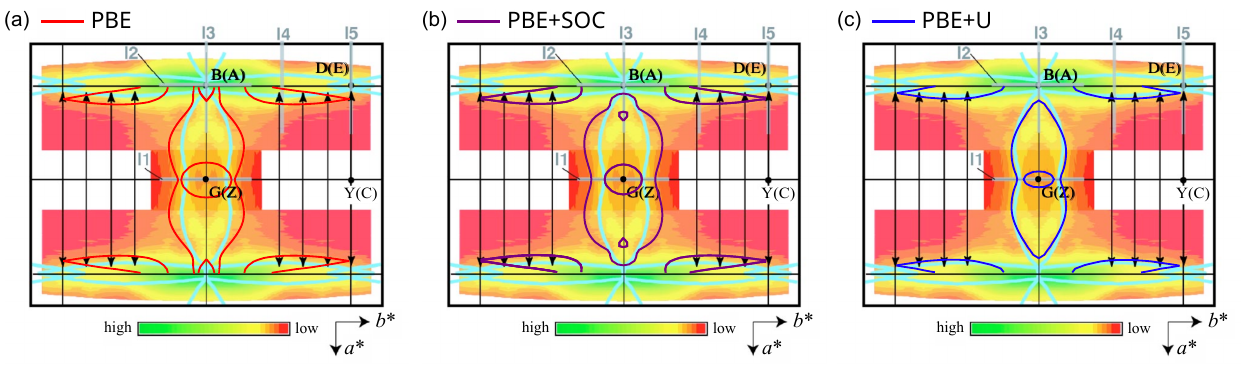}
\caption{Fermi surface of ZrTe$_3$ at $k_z$ = 0. The background color map shows the experimental ARPES intensity from Ref. \cite{Yokoya2005}. Black arrows, drawn in that study, indicate the q$_x$ component of the CDW vector. Cyan lines represent the theoretical FS from Ref. \cite{Stowe1998}. Our calculated FS cross sections are overlaid for comparison: (a) PBE (red), (b) PBE+SOC (purple), and (c) PBE+U applied to the Te atoms (dark blue).}
\label{Fig1}
\end{center}
\end{figure*}

Given the central role of Fermi surface nesting in this system, we first assess whether our high-temperature electronic structure calculations reproduce the experimental Fermi surface. As shown in Fig.~\ref{Fig1}(a), the Fermi surface obtained within PBE with fully relaxed atomic positions (red lines) fails to capture the ARPES measurements for $k_z$ = 0 of Ref.~[\onlinecite{Yokoya2005}] (color map). In particular, PBE yields overly delocalized and excessively hybridized Te $5p_x$ bands, producing quasi-1D Te(2)-Te(3) derived sheets that are incorrectly shaped. This deficiency originates in part from the relaxed internal coordinates, as PBE shortens interprism Te-Te distances (see Table \ref{tab:Te_distances}) and thereby enhances hopping amplitudes. Enforcing the experimental atomic positions within PBE improves agreement with ARPES data (see Fig. S3 \cite{supplemental_material}), similarly to earlier LDA LMTO-ASA results~\cite{Stowe1998} (cyan lines), but at the cost of generating large residual forces ($\sim$1 eV/\AA) that signal a structural instability. Moreover, the flattened oval pocket at the Brillouin zone center, which originates mainly from Te(1) 5$p_y$ states with some Zr $4d$ hybridization along the \textbf{b}-axis chains \cite{Stowe1998, Starowicz2007}, is also overly dispersive and incorrectly sized in PBE with relaxed coordinates. Despite the strong spin-orbit coupling (SOC) expected for heavy elements, its inclusion (purple lines in Fig.~\ref{Fig1}(b)) leaves the quasi-1D sheets essentially unchanged, affecting mainly Zr-derived bands such as the features around $\Gamma$ but failing to resolve the discrepancies with experiment.\\

A substantial improvement is obtained by including electron-electron interactions via on-site Hubbard corrections applied to the Te $5p$ states. Hubbard parameters have been computed from first principles using the linear-response approach of Timrov et al.~\cite{Timrov1,Timrov2}, yielding moderate values of $U_{\text{Te(1)}}=4.32$~eV, $U_{\text{Te(2)}}=4.38$~eV, and $U_{\text{Te(3)}}=4.44$~eV, indicative of intermediate correlations in the Te $5p$ orbitals. These values reflect the residual self-interaction in the Te $5p$ states within the employed exchange–correlation functional \cite{Cococcioni}. As shown by the dark-blue lines in Fig.~\ref{Fig1}(c), these corrections provide an essential enhancement to the electronic structure at the Fermi level and more generally, improve the band structure below it (see Fig. S4 \cite{supplemental_material}), together with optimized internal coordinates. The resulting Fermi surface shows excellent agreement with those reported in ARPES experiments, in particular reproducing more accurately the position and quasi-1D shape of the sheets along the B-D direction. Importantly, the effect of $U$ dominates over the choice of internal coordinates: PBE+U with PBE coordinates produces a better agreement with experiment than PBE using experimental coordinates (see Fig. S3 \cite{supplemental_material}). Thus, while both structural and correlation effects influence the character of the 5$p_x$ states along the Te(2)-Te(3) axis, the inclusion of an \textit{ab initio} Hubbard $U$ is the key factor in recovering the correct Fermi surface topology.\\

With the Fermi surface well characterized, we next assess whether its geometry can give rise to the CDW. We first compute the nesting function $\zeta(\mathbf{q})$, which is related to the imaginary part of the non-interacting susceptibility and identifies wavevectors \textbf{q} connecting regions of the Fermi surface:

\begin{equation} \label{eqn: nesting}
    \zeta(\mathbf{q}) = \frac{1}{N_\mathbf{k}} \sum_{nn'} \sum_{\mathbf{k}}^{1BZ} \delta(\epsilon_{n\mathbf{k}}) \delta(\epsilon_{n'\mathbf{k}+\mathbf{q}}) \,,
\end{equation}
where $\epsilon_{n\mathbf{k}}$ denotes the band energy measured with respect to the Fermi level. Peaks in $\zeta(\mathbf{q})$ indicate vectors connecting multiple portions of the Fermi surface, providing a quantitative measure of the nesting conditions relevant for CDW formation. \\

\begin{figure*}
\begin{center}
\includegraphics[width=2.0\columnwidth,draft=false]{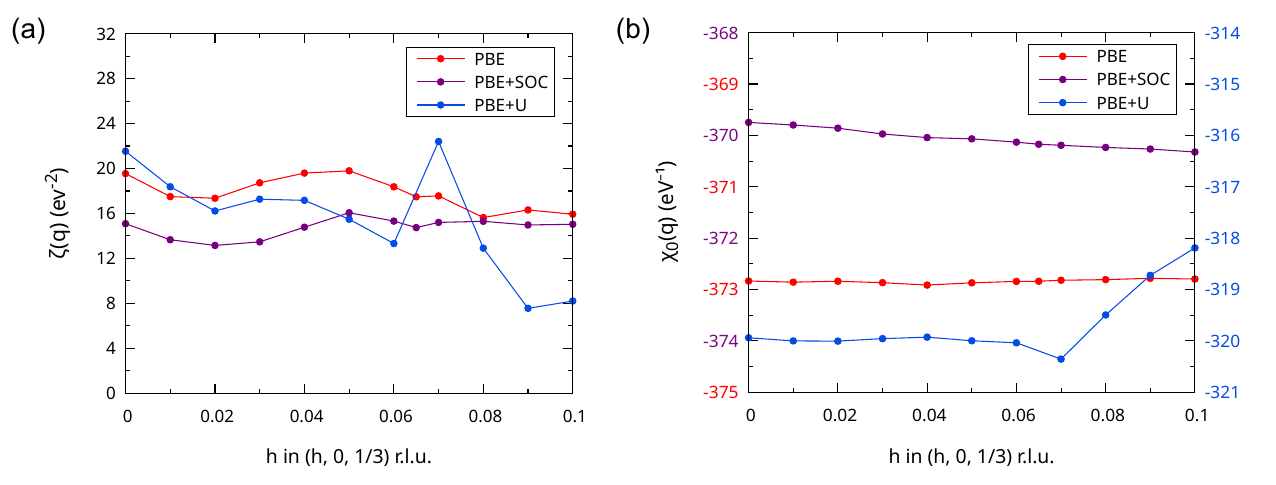}
\caption{Quantities related to the non-interacting static electronic susceptibility at different levels of theory (PBE: red; PBE+SOC: purple; PBE+U: blue). (a) Nesting function along ($h$, 0, 1/3) r.l.u for $h \in [0, \; 0.10]$. (b) Real part of the non-interacting susceptibility in the same momentum range. Note that different y-axis scales are used in panel (b) for PBE/PBE+SOC (left, red--purple) and for PBE+U (right, blue).}
\label{Fig2}
\end{center}
\end{figure*}

Figure \ref{Fig2}(a) shows the nesting function along the line ($h$, 0, 1/3) with $h \in [0, \; 0.10]$ in r.l.u., using the same color scheme as in Fig.~\ref{Fig1}. The PBE and PBE+SOC results exhibit no pronounced feature, indicating the absence of a well-defined nesting vector, in line with the fact that these calculations fail to reproduce the quasi-1D Fermi surface sheets responsible for nesting. The effect of SOC is therefore minimal and not relevant for the CDW formation. In contrast, the nesting function obtained within PBE+U displays a sharp and well-defined peak at h = 0.07 r.l.u, in excellent agreement with the experimental CDW wavevector. The intensity falls off rapidly for larger $h$ values, reflecting the highly localized nesting condition associated with the quasi-1D sheets. \\

Even if the nesting function $\zeta(\mathbf{q})$ provides information on the Fermi surface geometry, only a divergence in the real part of the susceptibility,  $\chi_0(\mathbf{q})$, can signal that an electronic instability is the underlying mechanism of the transition. This quantity is defined as:
\begin{equation}
    \chi_0(\mathbf{q}) = \frac{1}{N_\mathbf{k}} \mathcal{P} \sum_{nn'} \sum_{\mathbf{k}}^{1BZ} \frac{f_{n\mathbf{k}} - f_{n'\mathbf{k}+\mathbf{q}}}{\epsilon_{n\mathbf{k}} - \epsilon_{n'\mathbf{k}+\mathbf{q}}} \,,
    \label{eqn: chibare}
\end{equation}
where $\mathcal{P}$ denotes the principal value. This bare susceptibility enters in the definition of the phonon self-energy via Eqs. (22) and (23) in Ref. \cite{PhysRevB.82.165111}. While the contributions to $\zeta(\mathbf{q})$ come only from bands near the Fermi level, the real part of the non-interacting susceptibility may also involve deeper-lying bands. In practice, the momentum dependence mainly arises from pairs of filled and empty bands that are close in energy. Thus, \textbf{q} wavevectors that connect multiple regions of the Fermi surface and that produce peaks in $\zeta(\mathbf{q})$ may probably induce a localized softening on $\chi_0(\mathbf{q})$. Whether such contributions actually occur must nevertheless be checked explicitly.\\

In the present case, $\chi_0(\mathbf{q})$ was computed within the Wannier subspace. As shown in Fig. \ref{Fig2} (b), $\chi_0(\mathbf{q})$ in the HT-phase is essentially flat along the ($h$, 0, 1/3) direction for $h \in [0, \; 0.10]$ within both PBE and PBE+SOC functionals, a behavior that would typically signal negligible contributions from Fermi level states. However, when examined over the full Brillouin zone ($h \in [0, \; 1]$, Fig. S5 \cite{supplemental_material}), the values in this interval are moderately softened relative to those at larger $h$, indicating some degree of nesting for $|h| \lesssim 0.1$. This behavior originates from nearly parallel quasi-1D Fermi surface lamellae with weak out-of-plane dispersion. The absence of a pronounced \textbf{q}-dependent softening in $\chi_0(\mathbf{q})$ implies that this effect is broadly distributed in this $h$ range and does not single out any preferred wavevector. Together with the lack of pronounced peaks in $\zeta(\mathbf{q})$, these results indicate that no specific \textbf{q} wavevector connects a significant fraction of the Fermi surface within either PBE or PBE+SOC.\\

In the PBE+U calculations, we obtain the bare susceptibility by evaluating Eq. \ref{eqn: chibare} with the respective electronic structure. Within this framework, the overall magnitude of $\chi_0(\mathbf{q})$ decreases due to the downward shift of the low-energy bands, which do not contribute to the nesting and thus do not affect the analysis. For the smallest $h$ values, $\chi_0(\mathbf{q})$ is nearly constant, yet the extended Brillouin zone analysis reveals some degree of nesting without any single \textbf{q} wavevector dominating, as in PBE and PBE+SOC. Remarkably, at $\mathbf{q}_{\text{CDW}}$, the peak in the nesting function (Fig. \ref{Fig2} (a)) is translated as a weak softening in $\chi_0(\mathbf{q})$, which is absent in the other two functionals. At higher $h$ values, the absolute value of the real part of the susceptibility decreases, reflecting the reduced number of Fermi surface regions connected by these wavevectors once $U$ is included. \\

In conclusion, even though the visual distance between the parallel portions of the Fermi surface at $k_z$ = 0  does not differ significantly across the different computational schemes, the Fermi surface nesting is properly captured only when correlation effects on the Te $5p$ orbitals (which give rise to these parallel sections) are included. In the latter case, this nesting manifests itself as a localized peak in the nesting function and as a softening in the real part of the electronic susceptibility. It remains to be determined whether this kink alone is enough to drive the CDW transition or whether it acts in cooperation with the electron-phonon coupling. To address this question, we have performed DFPT calculations, which intrinsically include electron-phonon interactions and allow us to assess their combined effect.

\subsection{Vibrational properties}

\begin{figure}[b]
\begin{center}
\includegraphics[width=\columnwidth]{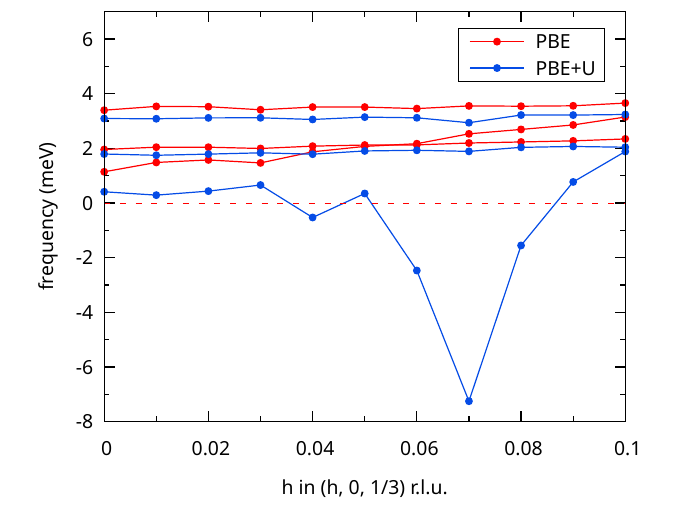}
\caption{Harmonic phonon spectra of ZrTe$_3$ calculated with PBE (red) and PBE+U (blue). Only acoustic modes are shown.}
\label{Fig3}
\end{center}
\end{figure}

\begin{figure*}[!t]
\begin{center}
\includegraphics[width=2.1\columnwidth,draft=false]{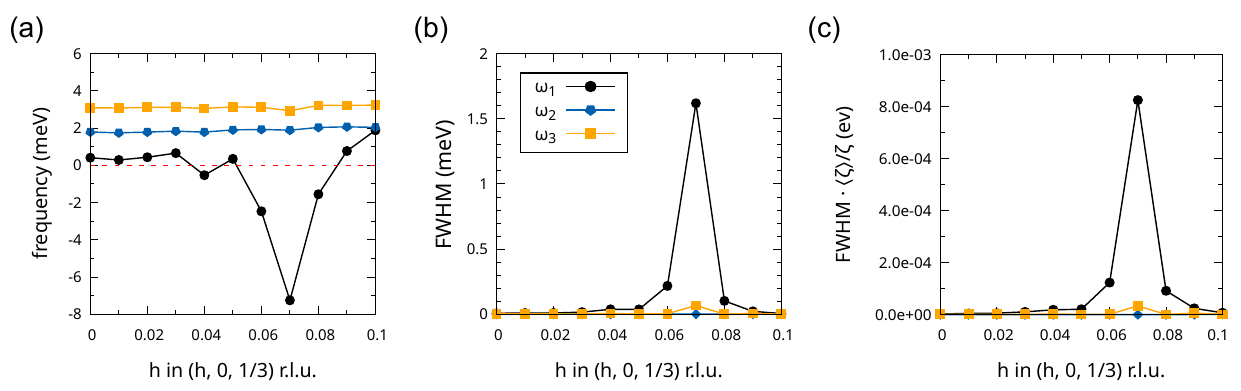}
\caption{Harmonic phonon spectra and electron-phonon interaction of calculated with PBE+U. (a) Acoustic phonon spectra along ($h$, 0, 1/3) r.l.u. for $h \in [0,\; 0.10]$. (b) Phonon linewidths (full width at half maximum) from electron--phonon coupling for the same modes and momentum range. (c) Ratio between the electron--phonon linewidths and the nesting function, calculated both with the same electronic parameters.}
\label{Fig4}
\end{center}
\end{figure*}

Fig. ~\ref{Fig3} shows the harmonic dispersion of the acoustic phonon branches along the ($h$, 0, 1/3) direction for $h \in [0,\; 0.10]$, comparing the results for PBE (red lines) and PBE+U (blue lines). Within PBE no imaginary phonon frequency is observed, and the system remains dynamically stable (also when including SOC, not shown). In contrast, once the on-site correlation effects between the Te $5p$ orbitals are included, the system develops a structural instability. In particular, an imaginary phonon frequency emerges at $h$ = 0.07 r.l.u., matching the experimentally established CDW wavevector. The fact that phonon softening comes out only when DFT+U correctly captures the Fermi surface topology clearly indicates that nesting plays a key role in determining \textbf{q}$_{\text{CDW}}$. Moreover, the softening extends over a narrow momentum range of approximately 0.04 r.l.u. ($\sim$0.04 \AA$^{-1}$) along this direction of the reciprocal space, a signature traditionally associated with Fermi surface nesting and in close agreement with the extent observed in IXS measurements \cite{Hoe09}. An explicit comparison between theory and experiment is presented in Fig. S7 of the Supplemental Material \cite{supplemental_material}.\\

From this point on, we consider only DFT+U in our calculations, as it is necessary to accurately capture the unstable phonon mode at \textbf{q}$_{\text{CDW}}$. We first analyze the contribution of the electron-phonon interaction to the phonon linewidth, which is given by the following formula for the full width at half maximum (FWHM):

\begin{equation} \label{eqn: lw}
\small
\gamma_{\mu}(\textbf{q})=\frac{4\pi w_{\mu}(\textbf{q})}{N_{\textbf{k}}}\sum_{nn'}\sum_{\textbf{k}}^{1BZ}|g^{\mu}_{n'\textbf{k}+\textbf{q},n\textbf{k}}|^2 \delta(\epsilon_{n'\textbf{k}+\textbf{q}})\delta(\epsilon_{n\textbf{k}}) \;,
\end{equation}
where $w_{\mu}(\textbf{q})$ is the harmonic phonon frequency and $g^{\mu}_{n'\mathbf{k}+\mathbf{q}, n\mathbf{k}}$ are the electron-phonon matrix elements. This expression mirrors that of the nesting function in Eq. \ref{eqn: nesting}, but weighted by the amplitude of the electron-phonon matrix elements-an aspect that we will exploit later to disentangle the microscopic origin of the CDW instability. \\

We computed the linewidths of the three acoustic modes along the same high-symmetry path as before, shown in Fig. \ref{Fig4} (b), with the corresponding harmonic phonon frequencies replotted for comparison in Fig. \ref{Fig4} (a). The lowest acoustic branch stands out, displaying both the strongest softening and the largest linewidth increase. In particular, at \textbf{q}$_{\text{CDW}}$, the localized harmonic instability of $\omega_1$ is accompanied by a sharp rise in its electron-phonon linewidth. These features appear to be even more pronounced than those in their electronic counterparts in Fig. \ref{Fig2} and provide an initial indication that the electron-phonon matrix elements exhibit a strong dependence on both mode and wavevector, a point that we examine in more detail below. \\

To clarify whether the electron-phonon interaction contributes to the softening of the CDW mode, or whether on the contrary the effect is purely electronic, we analyze the mode and momentum dependence of the electron-phonon matrix elements. A direct calculation of these coupling strengths is not feasible at the harmonic level when the system is dynamically unstable, since they scale as $w_{\mu}$(\textbf{q})$^{-1/2}$. However, a quantity proportional to the electron-phonon matrix elements can be obtained by evaluating the ratio between the electron-phonon linewidth in Eq. \ref{eqn: lw} and the nesting function in Eq. \ref{eqn: nesting} for each phonon branch. By further multiplying this ratio by the Brillouin zone average of the nesting function, $\langle \zeta \rangle_\textbf{q} = \mathrm{DOS}(E_F)^2$, i.e. considering ($\gamma_\mu(\mathbf{q}) / [\zeta(\mathbf{q}) / \langle \zeta \rangle_\textbf{q}]$), we obtain a normalized quantity having units of energy that directly reflects the momentum-dependent variations of the square of the electron-phonon coupling strength. We also note that $\gamma_\mu(\mathbf{q}) / [\zeta(\mathbf{q}) / \langle \zeta \rangle_\textbf{q}]$ does not depend on nesting. If the matrix elements were momentum independent, this normalized ratio would be flat for each of the modes. However, as shown in Fig. \ref{Fig4} (c), the ratios obtained for the acoustic branches display a pronounced mode and momentum dependence. Moreover, each of them closely follows the shape of the corresponding electron-phonon linewidth plotted to the left. This behavior indicates that the momentum dependence of the electron-phonon coupling outweighs that of the electronic nesting, and therefore also plays an important role in driving the CDW instability in this material. Indeed, the momentum dependence of the softest phonon branch essentially corresponds to the non-interacting electronic susceptibility of Fig. \ref{Fig2} (b) selectively amplified by the electron-phonon matrix elements at \textbf{q}$_{\text{CDW}}$, which ultimately drive the instability.

\subsection{Low symmetry CDW phase}

Finally, we determine the atomic structure of the low-symmetry CDW phase, which remains experimentally unresolved and has been proposed to be chiral. Displacing atoms along the eigendisplacement of the unstable phonon mode at $\mathbf{q}_{\mathrm{CDW}}$ within a $14 \times 1 \times 3$ supercell yields an initial energy gain of 0.13 meV per unit cell. Subsequent structural relaxation reduces the average force per atom from $\sim$0.016 eV/\AA\ to $\sim$0.005 eV/\AA\ and stabilizes the system by a total of 0.30 meV per unit cell. The relaxed structure (see Fig. S8 \cite{supplemental_material}) exhibits a symmetry reduction from $P2_1/m$ (No. 11) to $Pm$ (No. 6), retaining a mirror plane perpendicular to the $\mathbf{b}$ axis and thereby excluding a chiral CDW modulation.\\

\begin{figure}[!t]
\begin{center}
\includegraphics[width=1.0\columnwidth,draft=false]{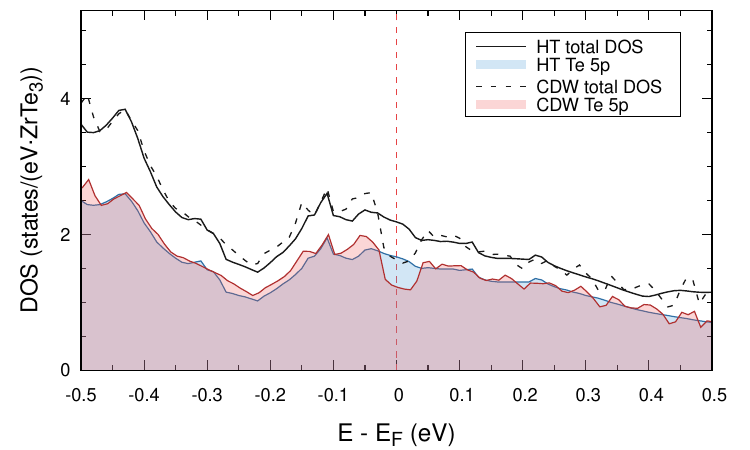}
\caption{Density of states around the Fermi energy for the high-temperature phase (solid black line) and for the low symmetry CDW phase (dashed black line). Te $5p$ contributions are indicated by lines with shaded areas: blue for the HT-phase and red for the CDW structure.}
\label{Fig5}
\end{center}
\end{figure}

Figure \ref{Fig5} shows the suppression of the density of states near the Fermi energy upon transitioning from the HT structure (solid black line) to the CDW phase (black dashed line). The red line with shaded area indicates that this spectral weight depletion originates predominantly from Te $5p$ states, consistent with ARPES measurements showing that the CDW gap opens on quasi-1D Fermi surface sheets with predominant Te $5p_x$ character. Furthermore, the calculated DOS closely recovers the experimentally observed spectral weight redistribution over $\sim$250 meV \cite{Yokoya2005}. Taken together, these results show that the obtained nonchiral CDW structure reproduces consistently the experimentally observed pseudogap formation.

\subsection{Conclusions}

Our results show that the charge density wave formation in ZrTe$_3$ arises from the cooperative action between Fermi surface geometry and electron-phonon coupling. The correct description of the electronic structure is essential, not only to identify the CDW wavevector but also to generate an instability in the harmonic phonon spectrum. In this regard, the modest enhancement of $\chi_0(\mathbf{q})$ at the critical wavevector originates from the proper matching between quasi-1D Fermi surface sheets, which is obtained only by including correlation effects on the Te $5p$ orbitals. The mode and momentum dependence of the electron-phonon coupling, however, plays the decisive role in determining the phonon mode responsible for the instability, i.e. other phonon modes do not soften, and in stabilizing the ordered state \cite{johannes2009}. Therefore, our work shows that even in quasi-one-dimensional compounds with chainlike structures resembling ideal Peierls chains, the selective action of the electron-phonon matrix elements is essential to fully understand the CDW formation and stabilization, highlighting the general importance of lattice dynamics in lowest dimensional materials. \\

\section{Acknowledgements}
The authors acknowledge M. Alkorta, M. Furci, G. Marini and S. Mocatti for technical assistance and valuable discussions.\\

\textbf{Funding information.} Funded by the European Union (ERC, DELIGHT, 101052708). Views and opinions expressed are however those of the authors only and do not necessarily reflect those of the European Union or the European Research Council. Neither the European Union nor the granting authority can be held responsible for them.

\section{Data Availability}
The data that support the findings of this article are openly available \cite{zenodo}.

\end{document}